\documentclass[smallextended]{svjour3}       % onecolumn (second format)
\smartqed  % flush right qed marks, e.g. at end of proof
\usepackage{graphicx,mathptmx,color,subfigure,amsmath,amssymb,bm}

\usepackage{cite}

\newcommand{\Tr}{\mathop{\rm Tr}}
\newcommand{\Z}{\mathbb{Z}}
\newcommand{\openone}{\leavevmode\hbox{\small1\normalsize\kern-.33em1}} 

\newcommand{\braket}[2]{\langle#1|#2\rangle}
\newcommand{\bra}[1]{\langle#1|}
\newcommand{\ket}[1]{|#1\rangle}

\journalname{Quantum Information Processing}

\begin{document}

\title{Tomography from collective measurements}



\author{A. Mu\~{n}oz$^{1,2}$ \and A. B. Klimov$^{^{1}}$  
 \and M.~Grassl$^{3}$ \and  L.~L.~S\'{a}nchez-Soto$^{3,4}$}


\institute{$^{1}$  Departamento de F\'{\i}sica, 
Universidad de Guadalajara, 44420~Guadalajara, 
Jalisco, Mexico \\
$^{2}$ Departamento de F\'{\i}sica, Universidad de Concepci\'on,
Casilla  160--C, Concepci\'on, Chile \\
$^{3}$ Max-Planck-Institut f\"ur die Physik des Lichts,
Staudtstra\ss e 2, 91058 Erlangen, Germany\\
$^{4}$ Departamento de \'Optica, Facultad de F\'{\i}sica, 
Universidad Complutense, 28040~Madrid,  Spain} 

\date{Received: date / Accepted: date}

\maketitle

\begin{abstract}
  We discuss the tomography of $N$-qubit states using
  collective measurements.  The method is exact for symmetric states,
  whereas for not completely symmetric states the information
  accessible can be arranged as a mixture of irreducible SU(2) blocks.
  For the fully symmetric sector, the reconstruction protocol can be
  reduced to projections onto a canonically chosen set of pure states.

\end{abstract}

\section{Introduction}

Continuous-variable tomography has been exhaustively explored, from
both  theoretical and experimental
viewpoints~\cite{Lvovsky:2009aa}. However, the corresponding problem
for discrete systems stands as challenge~\cite{lnp:2004uq}.  If we look
at the example of $N$ qubits, which will be our thread in this
paper, one has to make at least $2^{N}+1$ measurements in different
bases before to determine the state of \textit{a priori} unknown
system~\cite{Wootters:1989aa,Renes:2004aa,Lima:2011aa,Bent:2015aa}.
With such an exponential scaling, it is clear that only few-qubit
states can be reconstructed in a reasonable
time~\cite{Haffner:2005aa,Hou:2016aa}.

As a result, alternative techniques are called for. A wide class of
new protocols are explicitly targeted for particular types of
states. This includes states with low
rank~\cite{Gross:2010aa,Flammia:2012aa,Guta:2012aa,Riofrio:2017aa},
such as matrix product states
(MPS)~\cite{Cramer:2010aa,Baumgratz:2013aa}, or multiscale
entanglement renormalization ansatz (MERA)
states~\cite{Landon-Cardinal:2012aa}. The extra assumption of
permutationally invariance was also
examined~\cite{Ariano:2003aa,Toth:2010aa,
  Moroder:2012aa,Klimov:2013aa,Schwemmer:2014aa}, reducing the scaling
of the required setups to~$N^{3}$.

In the same spirit of simplicity, one may be tempted to examine the
case when one can extract only partial information from the system
under consideration.  This happens, e.g., in large multipartite
systems, wherein addressing individual particles turns out to be a
formi\-dable task.  Bose--Einstein condensates constitute an archetype
of this situation: only collective spin observables can be efficiently
measured through detection of the spontaneous emission correlation
functions~\cite{French:2003aa,Inguscio:2015aa}.

By assessing collective spin operators, one can only access the SU(2)
invariant subspaces appearing in the decomposition of the $N$-qubit
density matrix. The problem of partial state tomography appears thus
analogous to that of permutationally invariant states.

In the present work, we show that one can obtain an explicit partial
reconstruction for the $N$-qubit density matrix in terms of
average values of correlation functions of approximately $N^{3}$
collective spin operators. In other words, we propose to arrange
$O(N^{3})$ experimental data points inside SU(2) invariant
subspaces. As an illustration, we analyze the fidelity of the
reconstructed states for $2$ and $3$ qubits.  In addition, we
demonstrate that when the state belongs to the fully symmetric
(Dicke) subspace, the tomographic measurements reduce to rank-one
positive operator valued measurements (POVMs), and we find the
corresponding operational expansion.  As a bonus, we introduce a new
type of discrete special functions that might find further
applications in the analysis of $N$-qubit systems.

The paper is organized as follows. In Section~\ref{StandarTomo} we
briefly recall the principal aspects of discrete phase-space
distribution functions and of the standard tomographic scheme.  In
Section~\ref{Complete} we provide explicit expressions for the
permutationally invariant tomography for a $N$-qubit system, whereas
in Section~\ref{Canonical_P} an alternative scheme for fully symmetric
states is presented. Finally, Section~\ref{Conclusion} summarizes our
main results.

\section{Standard discrete tomography}
\label{StandarTomo}

For a system of $N$ qubits, the Hilbert space is the tensor product
$\mathbb{C}^{2} \otimes \cdots \otimes \mathbb{C}^{2} =
\mathbb{C}^{2N}$. The generators of the Pauli group $\mathcal{P}_{N}$
can be written as~\cite{Chuang:2000aa,Bjork:2008fk}
\begin{equation}
\hat{Z}_{\alpha} = 
\hat{\sigma}_{z}^{\alpha_{1}} \otimes \cdots \otimes 
\hat{\sigma}_{z}^{\alpha_{N}} \, ,
\qquad
\hat{X}_{\beta} = 
\hat{\sigma}_{x}^{\beta_{1}} \otimes \cdots \otimes
\hat{\sigma}_{x}^{\beta_{N}} \, ,
\end{equation}
so that they are labeled by $N$-tuples
$\alpha =(\alpha_{1},\ldots,\alpha_{N})$ and
$\beta=(\beta_{1},\ldots,\beta_{N})$, with
$\alpha_{j}, \beta_{j}\in \mathbb{Z}_{2}$. Here, $\hat{\sigma}_{z}$
and $\hat{\sigma}_{x}$ are the usual Pauli operators on the $i$th
qubit: $\hat{\sigma}_{z}=|0\rangle\langle 0|-|1\rangle \langle 1|$ and
$\hat{\sigma}_{x}=|0\rangle \langle 1|+|1\rangle \langle 0|$ in the
orthonormal computational basis $\{ |0 \rangle , | 1 \rangle \}$.

We next define the operators
\begin{equation}
  \hat{\Delta}^{(s)}( \alpha,\beta) =
\frac{1}{2^{N(s+3)/2}} \sum_{\gamma,\delta\in\Z_2^N}
 ( -1) ^{\alpha\delta +\beta\gamma +\gamma \delta (1-s)/2}
 \langle \xi |\hat{Z}_{\gamma }\hat{X}_{\delta }|\xi\rangle^{-s}
 \hat{Z}_{\gamma }\hat{X}_{\delta }\, , 
\label{p}
\end{equation}
where $s = \pm 1$.  From a physical perspective, the fiducial state
$|\xi \rangle $ can be chosen as a factorized symmetric (with
respect to particle permutations) state
$ \ket{\xi} = \otimes_{i=1}^{N} \ket{\xi}_{i}$
\begin{equation}
  \ket{\xi}_{i} = \frac{1}{\sqrt{1+|\xi|^{2}}}
  (\ket{0}_{i}+\xi\ket{1}_{i}) \, ,
  \label{fiducial}
\end{equation}
and $\xi =\frac{\sqrt{3}-1}{\sqrt{2}}e^{i\pi /4}$, which corresponds
to a spin coherent state determined by the normalized vector
$\mathbf{n}=(1,1,1)/\sqrt{3}$ on the Bloch sphere~\cite{Munoz:2012aa}.
The operators $\hat{\Delta}^{(s)}$ form a biorthogonal operator basis, namely
\begin{equation}
\Tr [ \hat{\Delta}^{(1)} ( \alpha ^{\prime },\beta ^{\prime }) \, 
\hat{\Delta}^{(-1)} ( \alpha,\beta ) ] =2^{N}\delta _{\alpha \alpha
^{\prime }}\delta _{\beta \beta ^{\prime }}.  
\label{OR1}
\end{equation}
In complete analogy with the continuous
case~\cite{Schroek:1996aa,QMPS:2005aa}, any operator
$\hat{A}$ acting on the Hilbert space $\mathbb{C}^{2^{N}}$
can be expanded in this basis as
\begin{equation}
\hat{A} = \sum_{\alpha,\beta\in\Z_2^N } 
Q_{A} ( \alpha,\beta ) \; \hat{\Delta}^{(1)} ( \alpha,\beta) \, ,  
\label{R1}
\end{equation}
where $Q_{A} ( \alpha,\beta ) =\Tr [ \hat{A} \; \Delta ^{(-1)} ( \alpha ,\beta)]$.

Actually, the kernel $\hat{\Delta}^{(-1)}(\alpha,\beta)$ can be
represented as a rank-one projector
\begin{equation}
  \hat{\Delta}^{(-1)} ( \alpha,\beta ) =
  \ket{\alpha,\beta} \bra{\alpha,\beta} \, .
  \label{D-1}
\end{equation}
Here, $\ket{\alpha,\beta}$ are discrete coherent states, constructed
as~\cite{Galetti:1996aa,Marchiolli:2007aa,Munoz:2009aa,
Klimov:2009aa,Klimov:2009bk}
\begin{equation}
  \ket{\alpha,\beta} = \exp[i  \chi( \alpha,\beta)] \;
 \hat{Z}_{\alpha}\hat{X}_{\beta }\ket{\xi} \, ,  
\label{cs}
\end{equation}
where $\exp[i \chi( \alpha,\beta)]$ is an appropriately chosen phase that
is irrelevant for our purposes here.  Up to normalization, the set of
projectors \eqref{D-1} forms an informationally complete
POVM~\cite{Prugovecki:1977fk,Busch:1989kx} with the choice
\eqref{fiducial} for the fiducial state. They  satisfy the
condition
\begin{equation}
\sum_{\alpha,\beta\in\Z_2^N} \ket{\alpha,\beta} \bra{\alpha,\beta} 
= 2^{N} \openone \, .
\end{equation}
In the single qubit case they form a SIC-POVM with four elements, and
for $N$ qubits, they correspond to tensor
products of the single-qubit SIC-POVM elements.  Equation~\eqref{R1}
can be thus interpreted as a tomographic reconstruction of the
operator $\hat{A}$ in terms of measured probabilities
$Q_{A}(\alpha,\beta) =
\bra{\alpha,\beta} \hat{A}\ket{\alpha,\beta}$. Moreover, 
$Q=\Tr[ \hat{A} \, \Delta ^{(-1)} ]$ and $P=\Tr [ \hat{A} \, \Delta^{(1)}]$
are discrete analogous of their continuous counterparts, defined in a
$2^{N}\times 2^{N}$ \ discrete phase
space~\cite{Galetti:1992aa,Ruzzi:2005aa}.  

\section{Tomography from collective measurements}
\label{Complete} 

The representation \eqref{R1} requires measuring the POVM \eqref{D-1}
with $2^{2N}$ elements. This provides a minimal complete tomography, but
it is extremely demanding for $N \gg 1$. 

As heralded in the Introduction, to circumvent this problem we restrict
ourselves to collective  measurements.  The information acquired from such
measurements does not allow to obtain complete information about the
state of the system: operators that are invariant under particle
permutations (collective operators) ``see'' only irreducible subspaces
appearing in the tensor decomposition of SU(2)$^{\otimes N}$.
Nonetheless, this still provides nontrivial information.

Symmetric operators $\hat{A}_{\text{sym}}$ on $N$ qubits are those
invariant with respect to particle permutations:
\begin{equation}
\hat{A}_{\text{sym}} = \hat{\Pi}_{ij}^{\dagger} \;
\hat{A}_{\text{sym}} \; \hat{\Pi}_{ij} 
\qquad
\forall i,j=1,\ldots,N,  \quad i\ne j \, ,
\end{equation} 
where $\hat{\Pi}_{ij}$ is the unitary operator that
swaps particles $i$ and $j$.  
The crucial observation for what follows
is that these operators possess a peculiar property: their
symbols $P_{A_{\text{sym}}}(\alpha,\beta)$ depend exclusively on the
Hamming weights~\cite{Montanaro:2009aa} of $\alpha$, 
$\beta$,  and  their binary sum $\alpha+\beta$; that is,
\begin{equation}
P_{A_{\text{sym}}} (\alpha,\beta) = 
P_{A_{\text{sym}}} \bigl( h(\alpha ), h(\beta), h(\alpha+\beta )\bigr)
\, ,
\end{equation}
with $h ( \kappa ) = |\{i\colon i=1,\ldots,N|\kappa_i\ne 0\} |$,
$0\leq h( \kappa)\leq N$. 

Therefore, the whole information about any symmetric measurement
$\langle \hat{A}_{\text{sym}}\rangle = \Tr(\hat{\rho}
\hat{A}_{\text{sym}})$ is conveniently conveyed in the projected
$\tilde{Q}$-function \cite{Klimov:2014aa,Gaeta:2016aa},
\begin{equation}
   \tilde{Q}_{\rho} ( m,n,k) = \sum_{\alpha,\beta \in\Z_2^N}
  Q_{\rho}( \alpha,\beta ) \;
  \delta _ {h( \alpha ) ,m} \, \delta_{h( \beta ) ,n} \, 
  \delta _{h( \alpha +\beta ) ,k} \, ,
  \label{Qmnk}
\end{equation}
since, as it immediately follows from \eqref{R1},
\begin{equation}
  \langle \hat{A}_{\text{sym}}\rangle = 
  \sum_{m,n=0}^{N}\sum_{k} 
  P_{A_{\text{sym}}}(m,n,k) \; \tilde{Q}_{\rho } ( m,n,k) \, ,  
  \label{f_av}
\end{equation}
the index $k$ running in steps of two:
$k= | m-n| , | m-n | +2, \ldots, \min(m+n,N,2N-m-n)$.

If $\tilde{Q}_{\rho }( m,n,k) $ is available from measurements,  
one can  lift it from the three-dimensional $(m,n,k)$ space into the
full $2^{N}\times 2^{N}$  discrete phase-space according to
\begin{equation}
Q_{\rho }^{\text{lifted}}( \alpha,\beta ) =
R_{mnk}^{-1} \sum_{m,n,k}
\delta _{h( \alpha ) ,m} \,\delta_{h( \beta ) ,n} \, \delta _{h(\alpha +\beta ) ,k}
\; \tilde{Q}_{\rho }( m,n,k) \, ,  \label{QR} 
\end{equation}
where
\begin{equation}
R_{mnk} = \! \! \! \sum_{\mu,\lambda\in\Z_2^N } \! \!
\delta _{h\left( \mu \right) ,m} 
\delta_{h\left( \lambda \right) ,n} \delta _{h\left( \mu +\lambda \right) ,k}
= \frac{N!}{\left( \frac{m+n-k}{2}\right) ! 
\left( \frac{2N-m-n-k}{2}\right)! 
\left( \frac{n-m+k}{2}\right) !\left( \frac{m-n+k}{2}\right) !}
\label{Rmnk} 
\end{equation}
is a normalization factor fixed by the number of binary tuples
$\lambda,\mu$ with Hamming weights
$\bigl(h(\lambda),h(\mu),h(\lambda+\mu)\bigr)=(m,n,k)$. 

The reconstruction \eqref{QR} of the $Q_{\rho}(\alpha,\beta)$ function
from the projected one $\tilde{Q}_{\rho}(m,n,k)$ is incomplete; i.e., the
map \eqref{Qmnk} is not faithful. The lifting \eqref{QR} is thus just
a way of organizing information obtained from
$\binom{N+3}{3}=(N+1)(N+2)(N+3)/6$ collective measurements,
corresponding to the total number of possible triplets $(m,n,k)$ of
Hamming weights, in a $2^{N}\times 2^{N}$ matrix.

By replacing $Q_{\rho }(\alpha,\beta)$ by
$Q_{\rho}^{\text{lifted}}(\alpha,\beta) $ in the reconstruction
\eqref{R1}, we get
\begin{equation}
\hat{\rho}_{\text{rec}} = \sum_{m,n,k} R_{mnk}^{-1} \;
Q_{\rho }( m,n,k) \; \hat{\Delta}^{(1) }(m,n,k)\, ,  
\label{rho_rec} 
\end{equation}
where the symmetric operators $\hat{\Delta}^{(\pm 1) }(m,n,k)$ 
can be jotted down as 
\begin{eqnarray}
\hat{\Delta}^{( \pm 1 ) }(m,n,k) & = & 
\sum_{\alpha,\beta\in\Z_2^N} 
\delta _{h( \alpha),m} \, \delta _{h(\beta),n} 
\delta_{h(\alpha+\beta),k} \; 
\hat{\Delta}^{(\pm1) }(\alpha,\beta) 
\nonumber \\
& = &  2^{-(3\pm1)N/2} \sum_{m',n',k'}
g_{mnk}( m',n',k') \, 3^{\pm (m'+n'+k')/4}\, \hat{F}_{m'n'k'} \, .
\end{eqnarray}
Here, $\hat{F}_{mnk}$ stands for the orthonormal set of operators (see
Appendix~\ref{Fmnk_base} for details)
\begin{equation}
\hat{F}_{mnk}= \sum_{\mu,\lambda\in\\Z_2^N } 
\delta _{h(\mu),m} \, \delta_{h(\lambda ),n} \, 
\delta_{h(\mu+\lambda) k}\; 
(-i)^{\mu \lambda }\hat{Z}_{\mu }\hat{X}_{\lambda } \, ,  
\label{F}
\end{equation}
and 
\begin{equation}
  g_{mnk} \bigl( h(\alpha), h(\beta), h(\alpha+\beta) \bigr) =
  \sum_{\gamma,\delta\in\Z_2^N}  (-1)^{\alpha\delta+\beta\gamma}\; 
  \delta_{h(\gamma),m} \, \delta_{h(\delta),n}\,
  \delta_{h(\gamma+\delta),k}
  \label{gmnk}
\end{equation}
are discrete functions, whose properties are explored in
Appendix~\ref{Specialfunctionf}. Observe that
\eqref{gmnk} is independent of the choice of $\alpha$ and $\beta$: any
other choice $\alpha'$ and $\beta'$ is related by permutations, which
can be applied to $\gamma$ and $\delta$ as well.
           
Because of the properties of $\hat{F}_{mnk}$, the reconstruction
\eqref{rho_rec} can be reduced to
\begin{equation}
\hat{\rho}_{\text{rec}} = \frac{1}{2^N}
\sum_{m,n,k} R_{mnk}^{-1} \langle\hat{F}_{mnk} \rangle \, 
\hat{F}_{mnk} \, ,
\label{rho_rec_f}
\end{equation}
which is an explicit function of $\binom{N+3}{3}$ expectation values
of collective operators. 

Note, in passing, that  the operators \eqref{F} can
be always expanded in terms of  collective spin operators. For
instance, by direct inspection one gets that in the simplest cases $n=0$
or $m=0$, $\hat{F}_{mnk}$ are diagonal in the computational basis:
\begin{equation}
\hat{F}_{000}= \openone\, 
\quad
\hat{F}_{101} = \hat{S}_{z},
\quad
\hat{F}_{202}= \frac{1}{2!} (\hat{S}_{z}^{2}-N \openone) ,
\quad
\hat{F}_{303}= \frac{1}{3!} [  \hat{S}_{z}^{3}-
(3N-2) \hat{S}_{z}]  \, , 
\end{equation}
where $\hat{S}_{j} = \sum_{i=1}^{N} \hat{\sigma}_{j}^{(i)}$.

By construction, $\hat{\rho}_{\text{rec}}$ is nonzero only inside
SU(2) invariant blocks, and coincides with the true density matrix in
the fully symmetric subspace.  In all the other blocks,
$\hat{\rho}_{\text{rec}}$ differs from the true value and,
in particular, the irreducible subspaces of the same dimension are
indistinguishable in $\hat{\rho}_{\text{rec}}$.

Let us illustrate the approach with a couple of basic examples.  
An arbitrary pure two-qubit state can be parametrized as
\begin{equation}
\ket{\psi} =\sin \theta \ket{\psi_\text{anti}} +
e^{i\beta} \cos\theta \ket{\psi_\text{sym}},
\end{equation}
where  $\ket{\psi_\text{sym}}$ and $\ket{\psi_\text{anti}}$ denote
states from the symmetric and antisymmetric subspaces
correspondingly, represented in  the computational basis as 
\begin{eqnarray}
\ket{\psi_\text{anti}} &=&
\frac{\ket{01} -\ket{10}}{\sqrt{2}}, \nonumber \\
& & \\
\ket{\psi_\text{sym}} &=&\sin \alpha _{1}\ket{00} +
e^{i\gamma _{1}}\cos \alpha_{1}\sin \alpha _{2}
\left( \frac{\ket{01} +\ket{10} }{\sqrt{2}}\right)
+ e^{i\gamma _{2}}\cos \alpha _{1}\cos \alpha _{2} \ket{11} \, , 
\nonumber 
\end{eqnarray}
$0\leq\alpha_1,\alpha_2\leq \pi/2$, $0\leq\gamma_1, \gamma_2<2\pi$.  
The reconstructed density matrix is the incoherent mixture 
\begin{equation}
  \hat{\rho}_{\text{rec}}= \sin ^{2} \theta  \; 
  \ket{\psi_\text{anti}}\bra{\psi_\text{anti}} +
  \cos^{2} \theta \;  \ket{\psi_\text{sym}} \bra{\psi_\text{sym}} \, .
\end{equation}
We quantify the accuracy of the reconstruction in terms of
the fidelity~\cite{Uhlmann:1976aa,Jozsa:1994aa}: $\mathcal{F} =
\bra{\psi}\hat{\rho}_{\text{rec}}\ket{\psi}$, which for this example reads
\begin{equation}
  \mathcal{F}=\frac{1}{4} [3+\cos(4\theta) ] \, ,
\end{equation}
so it depends only on the single parameter $\theta$ that determines
the projection onto the symmetric and antisymmetric subspaces,
respectively.  The minimum fidelity $\mathcal{F}=1/2$ corresponds to
the case when the subspaces have the same weight, whereas for states
in the completely symmetric or antisymmetric subspace, the
reconstruction is exact.

Our next example corresponds to an arbitrary pure three-qubit state,
which can be written as
\begin{equation}
  \ket{\psi}= \sin\theta\ket{\psi_{\text{sym}}} + 
  e^{i\beta}\cos\theta \, \sin\alpha \ket{\psi_{1}}
  +e^{i\gamma } \cos\theta \, \cos\alpha \ket{\psi_{2}},  
  \label{3PS}
\end{equation}
with $0 \leq \theta, \alpha \leq \pi/2$ and $0\leq\beta,\gamma<2\pi$.
$\ket{\psi_{\text{sym}}}$ is a state in the four-dimensional symmetric
subspace, whereas $\ket{\psi_1}$ and $\ket{\psi_2}$ are states in
SU(2)-irreducible two-dimensional subspaces. In the computational
basis, they are:
\begin{eqnarray}
\ket{\psi_{\text{sym}}} &= & \sin\theta_{1}\ket{000}+
  e^{i\beta_{1}}\cos\theta_{1}\sin\alpha_{1}
\frac{\ket{100} +\ket{010} +\ket{001} }{\sqrt{3}} \nonumber \\
& + & e^{i\gamma_{1}}\cos\theta_{1}\sin\alpha_{1}\sin\alpha_{2}
 \frac{\ket{110}+\ket{101} +\ket{011} }{\sqrt{3}}
+  e^{i\gamma_{2}}\cos\theta_{1}\cos\alpha_{1}\cos\alpha_{2}\ket{111},
\nonumber \\
\ket{\psi_{1}}& = & \sin\theta_{2}
 \frac{2\ket{100} -\ket{010} -\ket{001} }{\sqrt{6}}
 +e^{i\beta_{2}}\cos\theta_{2}\frac{\ket{101}-
  2\ket{011}+\ket{110}}{\sqrt{6}}, \nonumber \\
\ket{\psi_{2}}&= & \sin\theta_{3}
 \frac{\ket{001} -\ket{010} }{\sqrt{2}}
+e^{i\beta_{3}}\cos\theta_{3}\frac{\ket{101} -\ket{110} }{\sqrt{2}}.
\end{eqnarray}
The reconstructed density matrix is a mixed state, unless
$\ket{\psi}$ is in the symmetric subspace.  In
particular, the  blocks corresponding to
two-dimensional SU(2)-irreducible subspaces have the same form; viz,  
\begin{equation}
\hat{\rho}_{2}= \left(
\begin{array}{cc}
\sin^{2}\alpha\sin^{2}\theta_{2}+\cos^{2}\alpha\sin^{2}\theta_{3} & c_2 \\
c_2^*  &\cos^{2}\alpha\cos^{2}\theta_{3}+\cos^{2}\theta_{2}\sin^{2}\alpha
\end{array}
\right),
\end{equation}
where
$c_2=\tfrac{1}{2}[ e^{-i\beta_{2}}\sin^{2}\alpha\sin(2\theta_{2}) 
+e^{-i\beta_{3}}\cos^{2}\alpha\sin(2\theta_{3})]$ 
depends  on the parameters $\alpha$ and $\theta$ describing
the contributions from the three irreducible subspaces
$\mathcal{S}_{\text{sym}}$, $\mathcal{S}_{1}$, and $\mathcal{S}_{2}$
in   \eqref{3PS}. Thus, the
reconstructed density matrix  has the form
\begin{equation}
  \hat{\rho}_{\text{rec}}=\sin^{2}\theta 
\ket{\psi_{\text{sym}}}\bra{\psi_{\text{sym}}}
  \oplus\frac{\cos^{2}\theta}{2}\hat{\rho}_{2}
  \oplus\frac{\cos^{2}\theta}{2}\hat{\rho}_{2}.
\end{equation}
By averaging over the phases $\theta_2,\beta_2$ and $\theta_3,\beta_3$
that parameterize the states in the nonsymmetric irreducible subspaces
$\mathcal{S}_1$ and $\mathcal{S}_2$, we get the average fidelity that
determines the distribution between the SU(2)-irreducible subspaces:
\begin{equation}
  \bar{\mathcal{F}} =
\sin^2\theta +
\left [  \frac{3}{4} +\frac{1}{4}\cos^2(2\alpha)\right ] 
\cos^2\theta \, .
\end{equation}
The minimum $\bar{\mathcal{F}}_{\text{min}}=3/4$ corresponds to
the situation $\ket{\psi}=(\ket{\psi_{1}}+\ket{\psi_{2}})/\sqrt{2}$
($\alpha=\pi/4$, $\theta=0$) when the state is homogeneously
distributed between not completely symmetric subspaces
$\mathcal{S}_{1}$ and $\mathcal{S}_{2}$.  The maximum fidelity is
reached for symmetric states, $\ket{\psi}=\ket{\psi_{\text{sym}}}$.

\section{Symmetric overcomplete tomography: 
canonical projection}
\label{Canonical_P}

The outstanding case of fully symmetric (Dicke)
states~\cite{Dicke:1954aa} deserves special attention as they are
widely used in numerous applications (see,
e.g., ~\cite{Di:2005aa,Chiuri:2012aa,Apellaniz:2015aa}) and, in addition,
they are efficiently generated in the
laboratory~\cite{Duan:2003aa,Kiesel:2007aa,
  Wieczorek:2009aa,Prevedel:2009aa}.  For Dicke states, the
reconstruction \eqref{rho_rec_f} is exact, but requires $O(N^{3})$
measurements of collective operators, while the density matrix
contains at most $N^{2}+2N$ independent parameters.  Obviously, not
all such collective measurements are independent.  This redundancy can
be fixed by representing the reconstructed density matrix via rank-one
projectors.

For a fully symmetric density matrix it follows from \eqref{R1} that 
\begin{eqnarray}
  \hat{\rho}_{\text{sym}}  & = &  \hat{\Pi}_{\text{sym}}
\left(\sum_{\alpha,\beta\in\Z_2^N} 
Q_{\rho_{\text{sym}}}(\alpha,\beta) \;
\hat{\Delta}^{(1)}(\alpha,\beta)
\right)
\hat{\Pi}_{\text{sym}} \nonumber \\
&= &\sum_{\alpha,\beta\in\Z_2^N}
\Tr [\hat{\rho}_{\text{sym}} \;
\hat{\Delta}_{\text{sym}}^{(-1)}(\alpha,\beta) ] \;
\hat{\Delta}_{\text{sym}}^{\left( 1\right) }     \, ,  
\label{rho_sym}
\end{eqnarray}
with $\hat{\Delta}_{\text{sym}}^{(\pm 1)}(\alpha,\beta)=
\hat{\Pi}_{\text{sym}} \; \hat{\Delta}^{(\pm1)}(\alpha,\beta) \;
\hat{\Pi}_{\text{sym}}$ and  $\hat{\Pi}_{\text{sym}} = 
\sum_{\ell=0}^N \ket{\ell,N}\bra{\ell,N}$ is the projection onto the 
Dicke subspace $\{\ket{\ell,N}\colon\ell=0,\ldots,N\}$ of $N$ qubits.
 It is shown in Appendix~\ref{projectionk} that 
$\hat{\Delta}_{\text{sym}}^{(-1)}(\alpha,\beta) $ is a symmetric
function and actually it is a rank-one tensor
\begin{equation}
\hat{\Delta}_{\text{sym}}^{(-1)}( \alpha,\beta ) =
|\Psi _{h( \alpha) ,h( \beta ) ,h( \alpha +\beta ) }\rangle
\langle \Psi _{h( \alpha ) ,h( \beta ) ,h( \alpha+\beta ) }| \, .  
\label{D-1S}
\end{equation}
The unnormalized states
$\ket{\Psi_{h(\alpha),h(\beta),h(\alpha+\beta)}}$ have the following
expansion in the Dicke basis
\begin{equation}
  \ket{\Psi_{h(\alpha),h(\beta),h(\alpha+\beta)}} = 
\frac{1}{(1+|\xi|^{2})^{N/2}}
 \sum_{\ell=0}^N  \binom{N}{\ell}^{-1/2}
\psi_{\ell}\bigl(h(\alpha),h(\beta),h(\alpha+\beta);\xi\bigr) 
\ket{\ell,N} \, ,
\label{psi}
\end{equation}
$ \psi_{\ell}(h(\alpha),h(\beta),h(\alpha+\beta);\xi)$
being a discrete function discussed in Appendix~\ref{Specialfuntionpsi}. 

The operators \eqref{D-1S} form an informationally complete POVM
\begin{equation}
\sum_{m,n,k}N_{mnk}^{2} \; R_{mnk} \; 
\ket{\hat{\Psi}_{mnk}}\bra{\hat{\Psi}_{mnk}} =
2^{N} \hat{\Pi}_{\text{sym}} \, ,
\end{equation}
where $\ket{\hat{\Psi}_{mnk}} = N_{mnk}^{-1} \ket{\Psi _{mnk}}$ and 
\begin{equation}
N_{mnk}^{2}=\frac{1}{(1+|\xi|^{2})^{N}} 
\sum_{\ell=0}^N {\binom{N}{\ell}}^{{-1}} 
|\psi_{\ell}( m,n,k;\xi)|^{2} \, .
\end{equation}
For a given $(N+1)$-dimensional Dicke subspace, there are only $N$
different normalization factors $N_{mnk}$.

In terms of  the projection of $\hat{\Delta}_{\text{sym}}^{(1)}(\alpha,\beta)$, 
in Appendix~\ref{projectionk} we arrive at the  compact 
result
\begin{equation}
\hat{\rho}_{\text{sym}} = \sum_{m,n,k}
p_{mnk} \; R_{mnk} \; \hat{K}_{mnk}\, ,  
\label{rho_s}
\end{equation}
where  $p_{mnk}=\langle
\hat{\Psi}_{mnk}|\hat{\rho}_{\text{sym}}|\hat{\Psi}_{mnk}\rangle$ and
\begin{equation}
\hat{K}_{mnk}= 2^{-2N} 
\sum_{m',n,'k'} 3^{(m^{\prime}+n'+k')/4} \, 
i^{(k'-m'-n^{\prime})/2} \, g_{m'n'k'} ( m,n,k) \; \hat{A}_{m'n'k'} \, ,
\end{equation}
where $\hat{A}_{mnk}$ are also given in Appendix~\ref{projectionk}.

In this protocol, the total number of projections \eqref{D-1S}
required for reconstruction of symmetric states is
$\binom{N+3}{3}=(N+1)(N+2)(N+3)/6$. However, it immediately follows
from \eqref{rho_s} that the probabilities $p_{mnk}$ are not linearly
independent as they satisfy the  conditions
\begin{equation}
p_{m'n'k'}= \sum_{m,n,k} p_{mnk} \; \omega_{mnk}^{m'n'k'} \, ,  
\label{pj}
\end{equation}
where $\omega_{mnk}^{m'n'k'}= R_{mnk} \; \bra{\Psi_{m'n'k'}} 
\hat{K}_{mnk}\ket{\Psi_{m'n'k'}}$ and 
\begin{eqnarray}
  \omega_{mnk}^{m'n'k'} & = & \frac{R_{mnk}}{(1+|\xi|^{2}) ^{N}}
  \sum_{\ell,\ell'=0}^N \binom{N}{\ell}^{-1/2} \,
  \binom{N}{\ell'}^{-1/2}  \nonumber \\
& \times & 
  \psi_{\ell}(m,n,k;\xi) \; \psi_{\ell'}^{\ast}( m,n,k;\xi) \;
  f_{\ell\ell'}(m',k',n')\, .  \label{om}
\end{eqnarray}
These restrictions can be represented in a  matrix form
\begin{equation}
(\hat{\Omega} - \openone) \mathbf{p}  = 0 ,
\end{equation}
where $\mathbf{p}$ is the $\binom{N+3}{3}$-dimensional probability
vector and $\hat{\Omega}$ is an appropriately arranged matrix
\eqref{om}. We have numerically found that the rank of the matrix
$(\hat{\Omega} - \openone)$ is $N(N^{2}-1) /6$. Then, taking into
account that the probabilities also satisfy the normalization
condition $\Tr(\hat{\rho}_{\text{sym}}) =1$, we obtain that only
$N^{2}+2N$ projections are needed for the reconstruction of fully
symmetric states.

\section{Concluding remarks}
\label{Conclusion}

In short, we have proposed a tomographic protocol based on measuring
$\binom{N+3}{3}=(N+1)(N+2)(N+3)/6$ expectation values of collective
operators.  The advantage of the present approach with respect to
previously discussed (and experimentally verified) methods is
given by the explicit expressions \eqref{rho_rec_f} for the
reconstructed density matrix from experimental data.  In addition, we
have shown that restricting ourselves to fully symmetric states, the
tomographic protocol is reduced to projections from an overcomplete
set of pure states \eqref{psi}, which still allows to obtain an
explicit reconstruction expression \eqref{rho_s}.  Such a set of
states has been worked out from the first principles of state reconstruction in
an $2^{N}$-dimensional Hilbert space.

\begin{acknowledgements}
  This work is partially supported by the Grant 254127  of CONACyT
(Mexico). L.~L.~S.~S. acknowledges the support of the Spanish MINECO
(Grant FIS2015-67963-P).
\end{acknowledgements}

%%%%%%%%%%%%%%%%%%%%%%%%%%%%%%%%%%%%%%%%%%%%%%
\appendix

\section{Properties of the symmetric 
operators $\hat{F}_{mnk}$} 

\label{Fmnk_base}

The operators $\hat{F}_{mnk}$ can be expressed in terms of a special
discrete function.  Taking into account the action of the monomials
$\hat{Z}_{\alpha }\hat{X}_{\beta }$ on the computational basis states
$\{\ket{\kappa}\colon\kappa\in\Z_2^N\}$,
\begin{equation}
\hat{Z}_{\alpha }\ket{\kappa} =(-1)^{\alpha \kappa }\ket{\kappa} ,
\qquad \qquad 
\hat{X}_{\beta }\ket{\kappa} =\ket{\kappa +\beta} ,  
\label{ZX}
\end{equation}
we immediately obtain for the matrix elements
\begin{equation}
\bra{\delta}\hat{F}_{mnk}\ket{\gamma} = 
(-i)^{\frac{1}{2}(m+n-k)} \; \delta_{n,h(\delta+\gamma)} \,
f_{mk} \bigl(h(\delta),h(\gamma),h(\gamma+\delta)\bigr), 
\label{Fme}
\end{equation}
with
\begin{equation}
f_{mk}\bigl(h(\delta),h(\gamma),h(\gamma+\delta\bigr)) =
\sum_{\mu\in\Z_2^N}\delta_{h(\mu),m} \;
\delta_{h(\mu+\gamma+\delta),k} \; (-1)^{\mu\delta}. 
\label{fmk}
\end{equation}
The function$f_{mk}\bigl(h(\delta),h(\gamma),h(\gamma+\delta)\bigr)$ 
will be further analyzed below. 

By taking into account that $
\Tr (\hat{Z}_{\mu}\hat{X}_{\lambda}\hat{Z}_{\mu'}\hat{X}_{\lambda'})
=2^{N}(-1)^{\lambda\mu'}\delta_{\mu,\mu'}\delta_{\lambda,\lambda'}$,
we get
\begin{equation}
\Tr\left( \hat{F}_{mnk}\hat{F}_{m'n'k'}\right)
=2^{N}\delta_{m,m'} \; \delta_{n,n'} \; \delta_{k,k'}
\sum_{\mu,\lambda\in\Z_2^N}
\delta_{h(\mu),m} \; \delta_{h(\lambda),n} \; \delta_{h(\mu+\lambda),k} 
=2^{N} \; R_{mnk} \; \delta_{m,m'} \; \delta_{n,n'} \; \delta_{k,k'} \,,
\end{equation}
which shows the orthogonality used in the paper.

\section{Special functions}
\label{Specialfunction}

In this Appendix we discuss some relevant properties of the 
functions used in the derivation of our results. 

\subsection{Function $g_{mnk}$}

\label{Specialfunctionf} 

The discrete function \eqref{gmnk}
\begin{equation}
g_{mnk}\bigr(h(\mu),h(\lambda),h(\mu+\lambda)\bigl)  =
\sum_{\alpha,\beta\in\Z_2^N}(-1)^{\mu\beta+\lambda\alpha}
\delta_{h(\alpha),m} \; \delta_{h(\beta),n} \;
\delta_{h(\alpha+\beta),k} 
\end{equation}
can be represented in the integral form
\begin{eqnarray}
&&g_{mnk}\bigl( h( \mu ) ,h( \lambda ) ,h( \mu+\lambda ) \bigr)  
\nonumber \\
& & = \frac{1}{( 2\pi )^{3}} \oint\!\!\oint\!\!\oint 
\frac{d\omega_{1}d\omega_{2}d\omega_{3}}
{\omega_{1}^{m+1}\omega_{2}^{n+1}\omega_{3}^{k+1}}
( 1+\omega_{1}\omega_{2}+\omega_{1}\omega_{3}+
\omega_{2}\omega_{3})^{N-\frac{1}{2}[ h( \delta )  +
h(\gamma ) +h( \delta +\gamma ) ] } \nonumber \\
&& \times[1-\omega_{1}\omega_{2}-
\omega_{1}\omega_{3}+
\omega_{2}\omega_{3}]^{\frac{1}{2}[-h(\mu)+h(\lambda)+h(\mu+\lambda)]}\nonumber\\
&& \times[1-\omega_{1}\omega_{2}+
\omega_{1}\omega_{3}-
\omega_{2}\omega_{3}]^{\frac{1}{2}[h(\mu)-h(\lambda)+h(\mu+\lambda)]}\nonumber\\
&& \times [1+\omega_{1}\omega_{2}-
\omega_{1}\omega_{3}-
\omega_{2}\omega_{3}]^{\frac{1}{2}[h(\mu)+h(\lambda)-h(\mu+\lambda)]},
\end{eqnarray}
where we have used the following representation of the Kronecker
delta-function
\begin{equation}
\sum_{\kappa\in\Z_2^N }\delta_{h\left( \kappa \right) ,m} = 
\frac{1}{2\pi} \int_{0}^{2\pi }dx
e^{-ixm} \prod \limits_{i} \sum_{\kappa_{i}}e^{ix\kappa_{i}}.
\end{equation}
The above integrals can be easily computed, leading to a quite cumbersome
expression in terms of finite sums:
\begin{eqnarray}
& & g_{mnk}\bigl(h(\mu),h(\lambda),h(\mu+\lambda)\bigr)   = 
 \sum_{j_{1},\ldots,j_{10}} (-1)^{j_{3}+j_{4}+j_{6}+j_{5}-j_{1}-j_{7}} 
    \delta_{k,n+m-2j_{1}-2j_{5}-2j_{9}-2j_{10}} \nonumber \\ 
 & & =  \binom{j_{2}}{j_{6}} \binom{j_{3}}{j_{7}} \binom{j_{4}}{j_{8}} 
 \binom{j_{6}}{j_{10}}\binom{j_{7}}{j_{5}}\binom{j_{8}}{j_{1}}
\binom{N-\frac{1}{2}\left(h(\mu)+h(\lambda)+h(\mu+\lambda)\right)}
{n-\left(j_{1}+j_{2}+j_{3}+j_{4}+j_{5}+j_{9}+j_{10}-m\right)}
  \nonumber \\
& & \times
 \binom{\frac{1}{2}\left(-h\left(\mu\right)+h\left(\lambda\right)+h\left(\mu+\lambda\right)\right)}
{j_{2}} 
\binom{\frac{1}{2}\left(h(\mu)-h(\lambda)+h(\mu+\lambda)\right)}{j_{3}}
\binom{\frac{1}{2}\left(h(\mu)+h(\lambda)-h(\mu+\lambda)\right)}{j_{4}}
    \nonumber \\
& &\times
\binom{n-\left(j_{1}+j_{2}+j_{3}+j_{4}+j_{5}+j_{9}+j_{10}-m\right)}{m-\left(j_{6}+j_{7}+j_{8}\right)}
\binom{m-\left(j_{6}+j_{7}+j_{8}\right)}{j_{9}}.
\end{eqnarray}
They  satisfy the following dual orthogonality relations
\begin{eqnarray}
\sum_{m',n',k'}g_{mnk}(m',n',k') \; g_{m''n''k''}(m',n',k') \;
R_{m'n'k'}  
& = & 
2^{2N} \; R_{mnk} \; \delta_{m,m''} \; \delta_{n,n''} \; \delta_{k,k''},
      \nonumber \\
\sum_{m',n',k'} g_{m'n'k'}( m,n,k) \; g_{m'n'k'}(m'',n'',k'') \;  R_{m'n'k'}^{-1}
& = & 2^{2N} \; R_{mnk}^{-1} \; \delta_{m,m''} \; \delta_{n,n''} \; \delta_{k,k''}.
\end{eqnarray}

\subsection{Function $f_{mk}$}
Following a similar procedure, we can represent the function \eqref{fmk} in
the integral form,
\begin{eqnarray}
& &  f_{mk}\bigl(h(\delta),h(\gamma),h(\gamma+\delta)\bigr) 
= \frac{1}{(2\pi)^{2}}\oint\!\oint 
\frac{d\omega_{1}d\omega_{2}}{\omega_{1}^{m+1}\omega_{2}^{k+1}}
 (1+\omega_{1}\omega_{2})^{N-\frac{1}{2}[h(\delta)+h(\gamma)+h(\delta
    +\gamma)]}
 \nonumber\\
& & \times 
(\omega_{1}+\omega_{2})^{\frac{1}{2}[h(\gamma)-h(\delta)+h(\delta+\gamma)]}  
(\omega_{2}-\omega_{1})^{\frac{1}{2}[h(\delta)-h(\gamma)+h(\delta +\gamma)]}
(1-\omega_{1}\omega_{2})^{\frac{1}{2}[ h(\delta)+h(\gamma)-h(\delta +\gamma)] }.
\label{intF}
\end{eqnarray}
Computing the integral \eqref{intF} and rearranging the corresponding
sums of binomial coefficients we obtain
\begin{eqnarray}
& & f_{mk}\bigl(h(\delta),h(\gamma),h(\gamma+\delta)\bigr)   
=(-1)^{m}\binom{\frac{1}{2}
\left[h(\delta) +h(\gamma)-n\right]}{\frac{1}{2}(m-n+k)}
\binom{\frac{1}{2}\left[h(\delta)-h(\gamma)+n\right]}{\frac{1}{2}(m+n-k)}
 \nonumber \\
& & \delta_{\frac{1}{2}( m+n-k)   \in\Z}\; \delta_{\frac{1}{2}(m-n+k)\in\Z} 
\nonumber \\ 
& & \times\, _{2}F_{1}\left(\frac{h(\delta)+h(\gamma)+n}{2}-N,
-\frac{m-n+k}{2},1+\frac{h(\delta) +h(\gamma) -m-k}{2};-1\right)  
\nonumber\\
&& \times\,_{2}F_{1}\left(\frac{-h(\delta) +h(\gamma)  +n}{2},
-\frac{m+n-k}{2},1+\frac{h(\delta)-h(\gamma)-m+k}{2};-1\right), 
\label{g}
\end{eqnarray}
where $_{2}F_{1}$ is the Hypergeometric function. It is worth noting that 
$f_{mk}\bigl(h(\delta),h(\gamma),h(\gamma+\delta)\bigr)$ can
be obtained by a reduction from
$g_{mnk}\bigl(h(\delta),h(\gamma),h(\gamma+\delta)\bigr)$.

\subsection{Function $\psi_{\ell}$}
\label{Specialfuntionpsi}
The  function $\psi_{\ell}$ is defined as
\begin{equation}
\psi_{\ell} \bigl ( h(\alpha) , h( \beta), h( \alpha + \beta); \xi
\bigr ) = \sum_{\kappa \in \Z_{2}^{N}} \xi^{h(\kappa + \beta)} 
(-1)^{\alpha \kappa} \delta_{h ( \kappa), \ell}
\end{equation}
and it can be recast as
\begin{eqnarray}
&& \psi_{\ell}\bigr(h(\alpha), h(\beta),h(\alpha+\beta),\xi \bigl) = 
\xi^{h(\beta) }\sum_{\kappa \in\Z_2^N}
\xi^{h(\kappa)-2\sum_{i}\beta_{i}\kappa_{i}}(-1)^{\alpha \kappa }
\delta_{h(\kappa) ,\ell} \nonumber  \\
&& =\xi^{h(\beta) }\int \frac{d\omega }{\omega^{1+\ell}}
\prod_{i}^N\sum_{\kappa_{i}=0}^1
\xi^{\kappa_{i}-2\beta_{i}\kappa_{i}}(-1)^{\alpha_{i}\kappa_{i}}\omega^{\kappa_{i}} 
nonumber \\
& & =\xi^{h(\beta) }\int \frac{d\omega }{\omega^{1+\ell}}
(1+\xi \omega)^{N-\frac{1}{2}[ h(\alpha)+h(\beta)+h(\alpha+\beta) ] }
(1-\xi \omega)^{\frac{1}{2}[ h(\alpha)-h(\beta)+h(\alpha+\beta) ] } 
\nonumber \\
& &    \times 
( 1+\xi^{-1}\omega)^{\frac{1}{2}[-h(\alpha)+h(\beta)+h(\alpha+\beta) ]} 
(1-\xi^{-1}\omega)^{\frac{1}{2}[ h(\alpha)+h(\beta)-h(\alpha+\beta) ] },
\end{eqnarray}
which leads to the following expression in terms of finite sums
\begin{eqnarray}
&& \psi_{\ell} \bigl(h(\alpha),h(\beta),h(\alpha+\beta),\xi \bigr )
   \nonumber \\
& & =\xi^{l+h(\beta) }\sum_{j_{2},j_{3},j_{4}} 
(-1)^{j_{3}+j_{4}}\xi^{-2(j_{2}+j_{4})}
\binom{N-\frac{1}{2}[h(\alpha)+h(\beta)+h(\alpha+\beta)]}{l-j_{2}-j_{3}-j_{4}}
\nonumber \\
&& \times 
\binom{\frac{1}{2}[-h(\alpha)+h(\beta)+h(\alpha+\beta)]}{j_{2}} 
\binom{\frac{1}{2}[h(\alpha)-h(\beta)+h(\alpha+\beta)]}{j_{3}}
 \binom{\frac{1}{2}[h(\alpha)+h(\beta)-h(\alpha +\beta)]}{j_{4}} \, .
\nonumber \\
\end{eqnarray}

\section{Canonical projection}
\label{Cprojection}
In this Section we find projections of the kernels 
$\hat{\Delta}^{(\pm  1)}(\alpha,\beta)$  onto the Dicke subspace.  

\subsection{Projection of $\hat{\Delta}^{(-1)}(\alpha, \beta)$}
\label{projectionk}

Taking into account the representation of the Dicke states in the 
 logical basis,
\begin{equation}
\ket{\ell,N} =\frac{1}{\sqrt{\binom{N}{\ell}}}
\sum_{\substack{\kappa\in\Z_2^N\\h(\kappa)=\ell}}\ket{\kappa},  \label{dicke}
\end{equation}
we obtain 
\begin{equation}
\hat{\Delta}_{\text{sym}}^{(-1) }=
\sum_{\ell,\ell'}\frac{1}{\binom{N}{\ell}\binom{N}{\ell'}}
\sum_{\substack{\kappa,\kappa_{1}\in\Z_2^N\\h(\kappa)=\ell\\h(\kappa_1)=\ell}}
\sum_{\substack{\kappa',\kappa'_{1}\in\Z_2^N\\h(\kappa')=\ell'\\h(\kappa'_1)=\ell'}}
\ket{\kappa}\braket{\kappa_{1}}{\alpha,\beta}
\braket{\alpha,\beta}{\kappa'}\bra{\kappa_{1}'}, \label{delta_q_1}
\end{equation}
where the discrete coherent states $\ket{\alpha,\beta}$ are defined in
\eqref{cs}. Using the expansion of the fiducial state \eqref{fiducial}
in the logical basis
\begin{equation}
\ket{\xi} =\frac{1}{(1+|\xi|^{2})^{N/2}}
\sum_{\kappa\in\Z_n^2}\xi^{h(\kappa) }\ket{\kappa} ,
\end{equation}
we get
\begin{equation}
\braket{\kappa}{\alpha,\beta} = 
\frac{1}{(1+|\xi|^{2})^{N/2}}
\xi^{h(\kappa+\beta)}(-1)^{\alpha \kappa} \, .  \label{productos}
\end{equation}
By substituting \eqref{productos} into \eqref{delta_q_1}, we arrive at
\eqref{D-1S}.

The operator $\hat{\Delta}_{\text{sym}}^{(1)}(\alpha,\beta)$ can also be
expressed as
\begin{equation}
  \hat{\Delta}_{\text{sym}}^{(1)} (h(\alpha),h(\beta),h(\alpha+\beta))
 =\frac{1}{2^{2N}}\sum\limits_{m,n,k}\; 3^{(m+n+k)/4}i^{(k-m-n)/2} \;
g_{mnk}\bigl(h(\alpha),h(\beta),h(\alpha+\beta)\bigr) \;
\hat{A}_{mnk} \, ,  
\label{D1S}
\end{equation}
where $g_{mnk}\bigl(h(\alpha),h(\beta),h(\alpha+\beta)\bigr)$ is
defined in \eqref{gmnk}, and the matrix
elements of the operators $\hat{A}_{mnk}$ in the Dicke basis are
\begin{equation}
  \bra{\ell',N}\hat{A}_{mnk}\ket{\ell,N}= 
\frac{1}{\sqrt{\binom{N}{\ell}\binom{N}{\ell'}}}f_{\ell\ell'}(m,k,n) . \label{A}
\end{equation}
Finally, using the summation rule
\begin{equation}
\sum_{\alpha,\beta\in\Z_2^N}f(\alpha,\beta) = \sum_{m,n,k}
\sum_{\alpha,\beta\in\Z_2^N} \delta_{h(\alpha),m} \; 
\delta_{h(\beta),n} \; \delta_{h(\alpha+\beta),k} \; f(\alpha,\beta)\;,   
\label{suma}
\end{equation}
we get the explicit expression \eqref{rho_s}.

\subsection{Projection of monomials 
$\hat{\Pi}_{\text{sym}}\hat{Z}_{\alpha }\hat{X}_{\beta }
\hat{\Pi}_{\text{sym}}$}

\label{projectionm}

It follows immediately from \eqref{ZX} that the matrix elements of the
monomial $\hat{Z}_{\alpha }\hat{X}_{\beta }$ in the Dicke basis
\eqref{dicke} have the form
\begin{equation}
\bra{\ell,N}\hat{\Pi}_{\text{sym}}\hat{Z}_{\alpha }\hat{X}_{\beta} 
\hat{\Pi}_{\text{sym}}\ket{\ell',N} = 
\frac{1}{\sqrt{\binom{N}{\ell}\binom{N}{\ell'}}}
\sum_{\mu\in\Z_2^N}(-1)^{\alpha \mu }\delta_{h(\mu),\ell}
\delta_{h(\mu+\beta),\ell'} = 
\frac{f_{\ell\ell'}\bigl(h(\alpha),h(\alpha+\beta),h(\beta)\bigr)}
{\sqrt{\binom{N}{\ell}\binom{N}{\ell'}}},
\end{equation}
where the function $f_{\ell\ell'}$ is defined in \eqref{fmk}.

%\bibliographystyle{spphys}       
%\bibliography{ColTom}   % name your BibTeX data base
%\input{QINP_v1.bbl}

\end{document}